\documentclass[aps,prb,twocolumn,showpacs,superscriptaddress,floatfix]{revtex4}

\usepackage{graphicx}
\usepackage{amssymb}
\usepackage{amsmath}
\usepackage{amsfonts}
\usepackage{amsthm}
\usepackage{verbatim}
\usepackage{bbm}
\usepackage{subfigure}

\newfont{\BBB}{msbm10 scaled\magstephalf}

\newcommand{\N}{\mathbb{N}}
\newcommand{\Z}{\mathbb{Z}}

\newcommand{\R}{\mathbb{R}}
\newcommand{\C}{\mathbb{C}}

\newcommand{\te}{\theta}
\newcommand{\sd}{\rtimes}
\newcommand{\Rcal}{\mathcal{T}_r}

\newcommand{\Acal}{\mathcal{A}}
\newcommand{\Ucal}{\mathcal{U}}

\newcommand{\Lcal}{\mathcal{T}_l}

\newcommand{\Orb}{\mathcal{O}}
\newcommand{\braid}{\mathcal{R}}

\begin{document}
\title{Defect mediated melting and
the breaking of quantum double symmetries}
\author{F.A.~Bais}
\affiliation{ Institute for Theoretical Physics, Valckenierstraat
65, 1018 XE Amsterdam, The Netherlands}
\author{C.J.M.~Mathy}
\affiliation{Department of Physics, Princeton University, Jadwin
Hall, Princeton, NJ 08544, United States} 
\date{January 25, 2006}

\begin{abstract}
\noindent In this paper, we apply the method of breaking quantum
double symmetries \cite{Bais:2002pb,Bais:2002ny,bmbreaking:2006} to
some cases of defect mediated melting. The formalism allows for a
systematic classification of possible defect condensates and the subsequent
confinement and/or liberation of other degrees of freedom. We also
show that the breaking of a double symmetry may well involve a
(partial) restoration of an original symmetry. A detailed
analysis of a number of simple but representative examples is given,
where we focus on systems with global internal and external (space)
symmetries. We start by rephrasing some of the well known cases
involving an Abelian defect condensate, such as the
Kosterlitz-Thouless transition and one-dimensional melting, in our
language. Then we proceed to the non-Abelian case of a hexagonal
crystal, where the {\it hexatic phase} \cite{Mullen,Nelson} is
realized if translational defects condense in a particular
rotationally invariant state. Other conceivable phases are also
described in our framework.
\end{abstract}

\maketitle \footnotetext[1]{bais@science.uva.nl}
\footnotetext[2]{cjmmathy@princeton.edu}

\section{ Introduction}
In two-dimensional (quantum) physics certain quantum groups and
Hopf algebras play an important role because these extended
symmetries allow for a treatment of ordinary and topological
quantum numbers on equal footing, whereas the natural appearance
of an R-matrix accounts for the topological interactions in the
system \cite{Bais:1991pe}. This means that both topological
defects and ordinary excitations (such as Goldstone modes) may
appear in the same representation of the underlying - often hidden
- Hopf algebra $\Acal$, even in non-Abelian situations where the
mutual dependence of these dual quantum numbers would otherwise be
untractable. For a brief, tailor made introduction to Hopf
algebras and the notational conventions we adopt in this paper, we
refer to Appendix A of a related paper \cite{bmbreaking:2006}, or
to a more general review \cite{dwpb1995}. These extended symmetry
concepts  have found interesting applications in the domain of
Quantum Hall liquids \cite{Slingerland:2001ea}, crystals and
liquid crystals \cite{mbnematic:2006}, (2+1)-dimensional gravity
\cite{Bais:1998yn}, and potentially also to rotating Bose-Einstein
condensates \cite{rlsr:2003,clrs:2005} and other systems described
by conformal field theory .

\begin{figure}[t]
\begin{center}
\includegraphics[scale=0.45]{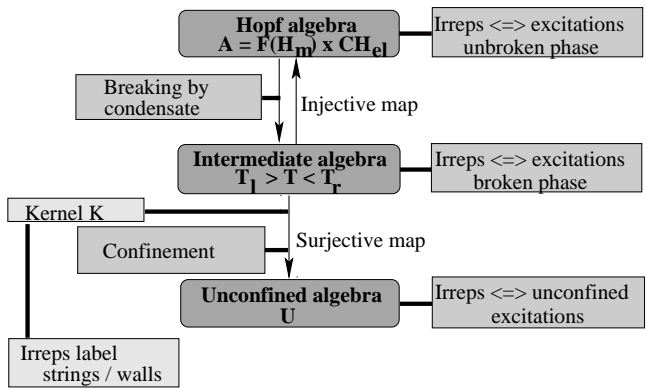}
\end{center}
\caption{A schematic of the quantum symmetry breaking formalism.
One has to distinguish three levels: The unbroken Hopf algebra
$D(H)$, the intermediate algebra $\Rcal$ and the effective
unconfined algebra $\Ucal$. This scheme should be compared with
the simple scheme of ordinary symmetry breaking where one only has
to distinguish between the unbroken  group $G$ and some residual symmetry
group $H \subset G$.} \label{quantumbreaking}
\end{figure}

Once this type of (hidden) symmetry was identified, a formalism
for the breaking of quantum double symmetries was proposed
\cite{Bais:2002pb,Bais:2002ny}, by assuming the formation of
condensates of ordinary (electric), defect (magnetic), or mixed
(dyonic) type. As was expected, the ordinary condensates reproduce the
conventional theory of symmetry breaking, though the systematic
analysis of the subsequent confinement of topological degrees of
freedom using the braid group is not standard.  In the papers just
mentioned it was shown that the usual picture of symmetry breaking
had to be augmented with significant novel ingredients. In the
case of Hopf double symmetry breaking we assume the condensate to
be represented by a fixed vector $|\phi_0\rangle$ in some
nontrivial representation $\Pi^A_\alpha$ of the Hopf algebra
$\Acal$. This leads to the definition of an intermediate algebra
$\mathcal{T}$ as the suitably defined stabilizer
subalgebra of the vacuum state: $\mathcal{T}$ was originally
defined as the largest Hopf subalgebra of $\mathcal{A}$ whose
elements $a$ satisfy
\begin{equation}
\Pi^A_\alpha(a)\mid \phi_0 \rangle = \varepsilon(a)\mid \phi_0
\rangle \qquad\forall \quad a \in \mathcal{T}\;,
\label{criterion1}
\end{equation}
where $\varepsilon(a)$ denotes the counit or trivial representation
of the Hopf algebra in question. This analysis is further
generalized in a related paper\cite{bmbreaking:2006}, where it is
argued that sometimes more general (non-Hopf) algebras play an
essential role. We then have to distinguish between a right or a
left intermediate algebra, $\mathcal{T}_r$ and $\mathcal{T}_l$
respectively, which are all related in the sense that $\mathcal{T}
\subseteq \mathcal{T}_r \cap \mathcal{T}_l$. The complication that
arises at this "intermediate" level is that certain representations
of $\mathcal{T}$ may have nontrivial braiding properties with the
vacuum state. This in turn means that the vacuum cannot be single
valued around a particle carrying that representation. If this
happens to be the case, it has the physical implication that such
particles (representations) have to be confined. The low energy
theory of non-confined degrees of freedom is then characterized by
yet a different algebra called $\Ucal$. So the breaking of Hopf
symmetries involves three algebras: the unbroken algebra $\Acal$,
the intermediate algebra (be it $\mathcal{T}$, $\Rcal$ or $\Lcal$),
and the unconfined algebra $\Ucal$. The generic breaking scheme is
depicted in Figure \ref{quantumbreaking}.
With this formalism of Hopf symmetry breaking available it is
important to take a closer look at suitable physical situations to
check whether the theory produces correct and useful results. This
is the subject of the remainder of this paper. Some applications
to non-Abelian nematic crystals are the subject of another paper
appearing in parallel \cite{mbnematic:2006}.

\section{{Abelian defect condensates}}
In this section we will treat some examples of ``Abelian''
condensates. The rather well known examples of Abelian defect
condensates are interesting because they cause confinement but at
the same time an analysis of the unconfined particles reveals that
a so called symmetry restoration takes place. In other words Hopf
symmetry breaking by defects may lead to the restoration of
ordinary symmetries.
\subsection{Kosterlitz-Thouless revisited}
The most well known example of a defect condensate is the one that
mediates the Kosterlitz-Thouless transition in the 2-dimensional
XY model \cite{KT}. In the isotropic phase, the internal symmetry
group is $SO(2)$. Representations of $SO(2)$ are labelled by an integer:
$n\in\Z $. For reasons that will become clear shortly, we also
introduce particles that transform under projective unitary
representations of $SO(2)$, which is equivalent to saying that
they transform under faithful representations of the universal
covering group $\R$ of $SO(2)$. These $\R$ representations are
labelled by a real number $\lambda\in\R$, giving the irrep
$\rho_{\lambda}$:
\begin{equation}
\label{eq:r-irreps}\rho_{\lambda}(x) = e^{i \lambda x}\;, \; x \in
\R.
\end{equation}
Thinking of $\lambda$ as labelling a spin degree of freedom one
has that the particle is bosonic if $\lambda\in\Z$, because
$\rho_{\lambda}(2\pi)=1$. If $\lambda\in\Z+\frac{1}{2}$, the irrep
is fermionic, because $\rho_{\lambda}(2\pi)=-1$. For any other
value of $\lambda$, the irrep would be called an Abelian anyon.

In the low temperature phase of our system, the $SO(2)$ symmetry is completely
broken by a bosonic ($\lambda=1$) condensate implying that  the
covering group $\R$ is broken to $\Z$. The residual symmetry
transformations, labelled by $n\in \Z$, correspond to rotations by $2\pi n$. The
Hopf symmetry in this phase is the \emph{quantum double} of $\Z$
denoted by $D(\Z)$, because the defects are labelled by an integer
$n$, as a consequence of the fact that
$\Pi_1(\R/\Z)=\Pi_0(\Z)=\Z$. The electric irreps in this phase are
labelled by $\mu\in [0,2\pi]$, corresponding to the irreps of
$\Z$. However, we can still include all the electric irreps of the
isotropic theory. In other words, we can restrict the irreps of
$\R$ to $\Z$, giving representations of $\Z$, some of which are
equivalent. We like to take all the electric irreps along, because
we want to see which irreps of the isotropic phase survive
unconfined in the various phases. Note that the Goldstone boson
transforms trivially under $\Z$, because $2\pi$ rotations don't
affect it. So in this phase we have the representations
$\Pi^n_\mu$ of $D(\Z)$, where
\begin{equation}\label{DZ}
D(\Z) = F(\Z)\times \C\Z .
\end{equation}
If we heat the system in this $D(\Z)$ phase, the
Kosterlitz-Thouless transition takes place corresponding to the
formation of a condensate of the defect $|\phi_0\rangle =
|n=1\rangle$. To see what happens one first has to determine the
intermediate algebra $\mathcal{T}$ and then use the braid
relations to determine which representations are confined to
obtain the effective unconfined algebra $\mathcal{U}$. The present
case corresponds to a single defect condensate and the general
structure of $\mathcal{T}$ and $\Ucal$ have been given elsewhere
\cite{bmbreaking:2006}. The outcome is
\begin{eqnarray}
&& \Rcal = F(H/(g_A))\times\C N_A \label{eq:mqdTsingdef}        \\
&& \Ucal = F(N_A/(g_A))\times\C N_A/(g_A).
\end{eqnarray}
where $(g_A)$ is the group generated by the group element $g_A$
corresponding to the defect in the condensate, and this group
$(g_A)$ is a normal subgroup of $N_A$, which is the normalizer in
$H$ of the element $g_A$. In the case we are considering, $(g_A)=N_A$, so for $\Ucal$
we obtain the rather trivial result:
\begin{equation}
\mathcal{U} = D(\Z/\Z)=D(e),\nonumber
\end{equation}
from which we conclude that there are no surviving defects in the
unconfined phase and also that from the set electric reps of
$H=\mathbb{Z}$ (i.e. $\mu\in [0,2\pi]$) non are left.

But what about the original set of spin representations
(\ref{eq:r-irreps})? Considering the braiding of a representation
with spin $\lambda$ with a defect $|n\rangle$, we obtain
that{\footnote{Note that $\braid$ involves a counterclockwise
"half braid", completely moving around is achieved by acting with
$\braid^2$. Braiding of a defect and a representation results in
just applying the group element corresponding to the defect on the
given representation. }
\begin{equation}
\braid^2 \cdot  |n\rangle |\lambda\rangle = e^{i 2\pi n
\lambda}|n\rangle |\lambda\rangle .
\end{equation}
The only irreps of $\R$ that braid trivially with the condensate
$|1\rangle$ are those labelled by an integer $m\in\Z$:
\begin{equation}
\rho_m(x)=e^{i2\pi m x}\;,\; x\in\R.\nonumber
\end{equation}
So the residual theory has no defects, and the unconfined electric
irreps are labelled by $n\in\Z$, in other words the fractional
charges are confined. A theory with no defects and charges
labelled by $n\in\Z$ is an $SO(2)$ theory.  This argument suggests
that \emph{the symmetry has been restored to a full $SO(2)$}, i.e.
the original $\R$ theory has been compactified, and the ``spin'' is
now quantized.

In general, condensing the $|n\rangle$ defect gives a theory with
$\Z_n$ defects (defects defined modulo n), and the electric irreps
are labelled by $\frac{k}{n},\;k\in\Z$, meaning that the symmetry
is restored to $SO(2)/\Z_n$.

\subsection{1-Dimensional melting}

In the previous example we studied a phase with internal symmetry,
we now turn to a similar example with a spatial translational
symmetry. Consider a one-dimensional system, say defined along the
x-axis. In the high energy phase, the external (space) symmetry
group of the system is the translation group $\R$. As in the
previous example, the irreducible unitary representations (irreps)
$\rho_{\lambda}$ of the external symmetry group $\R$ are again
labelled by $\lambda \in \R$ (see (\ref{eq:r-irreps})). We now break
this $\R$ symmetry to $\Z$, by the formation of a one dimensional
crystal, i.e. a 1-dimensional lattice. The lattice sites are at
positions $x=\ldots, -2a, -a, 0, a, 2a, \ldots$ We can label the
lattice sites by elements of $\Z$, where  $m\in\Z$ corresponds to
the lattice site at position $x=m a$.

The irreps $\Omega_{\mu}$ of the residual external symmetry group
$\Z$ are labelled by $\mu \in [0,2\pi/a)$:
\begin{equation}
\Omega_{\mu}(m) = e^{i\mu m a}, \; \; m\in\Z .
\end{equation}
$\mu$ is the \emph{momentum} of the irrep $\Omega_{\mu}$. In the
broken phase with broken continuous translation invariance
\emph{phonons} appear as Goldstone modes. These phonons have momenta
$k \in \R$. Furthermore the broken phase is also characterized by
possible defects (dislocations) labelled by an integer $n$ because
$\Pi_1(\R/\Z) = \Pi_0(\Z)=\Z$.

Now imagine that we condense the defect $|n\rangle$. If we now
consider the braiding of a phonon with momentum $k$ with this
defect, we obtain again that
\begin{equation}
\braid^2 \cdot  |n\rangle |k\rangle = e^{ikna} |n\rangle |k\rangle
.
\end{equation}
This braiding is trivial if and only if $e^{ikna} = 1$, which is
the case if k is an integer multiple of $\frac{2\pi}{na}$. Since
$k\in B$, we have $n$ phonons that braid trivially with the
condensate, namely the phonons $|k\rangle$ with momentum $k=0,
\frac{2\pi}{na}, 2 \frac{2\pi}{na}, \ldots, (n-1)
\frac{2\pi}{na}$. The defects $|m\rangle, m\in\Z$ are unconfined,
but they are now defined modulo $|n\rangle$, since $|n\rangle$ is
condensed.

Thus the condensation of $|n\rangle$ has led to a theory with $\Z_n$
defects, and $\Z_n$ phonons and we end up with an unconfined algebra
$ \mathcal{U}=D(\Z_n)$.

In particular, if we take $n=1$, i.e. we condense the $|1\rangle$
defect, we obtain a phase with trivial symmetry, in which there
are no defects, and all phonons are confined.

Let us now discuss melting. If we condense the defect $|1\rangle$,
then the unconfined modes are those with momentum $k=0,
\frac{2\pi}{a}, 2 \frac{2\pi}{a}, 3 \frac{2\pi}{a}, \ldots$. So all
phonons are confined (since a phonon cannot have exactly zero
crystal momentum), and the unconfined translational irreps have
momentum $k \in \{\frac{2\pi}{a}n,n\in\Z\}$. Thus we have a theory
with momentum modes labelled by $n\in\Z$, and no defects. This is
obviously a theory with a periodic translational $U(1)$ symmetry.
This is exactly the same argument as the one for the
Kosterlitz-Thouless phase, except that there the Goldstone modes
were not confined, yet the symmetry was restored to $SO(2)$. In this
theory one could of course also consider situations in which various
mutually compatible condensates (that trivially braid with each
other) occur simultaneously, leading to yet other residual
symmetries  and excitations.

\subsection{A defect condensate in a  2-dimensional smectic}

We may extend the previous reasoning to a particular 2-dimensional
system, where we consider only translational degrees of freedom.
In the high energy phase, the (translational) symmetry of the
system is $\R^2$. The irreps $\rho_{\vec{k}}$ of $\R^2$ are
labelled by $\vec{k}=(k_x,k_y)\in\R^2$:
\begin{equation}
\rho_{\vec{k}}(x,y) = e^{i(k_x x + k_y y)}.
\end{equation}
We break this symmetry to $\R\times\Z$, by condensing the mode
$\rho_{(0,a)}$. The system is then composed of a set of lines
parallel to the x-axis. In the y-direction, we have the same
situation as above, i.e. the symmetry is broken from $\R$ to $\Z$
in the y-direction. In the x-direction, the symmetry is still
unbroken, so we can neglect the x-direction. Let us be more
precise about this. The phonons, as Goldstone modes of the broken
translations in the y-direction have a momentum that points in the
y-direction so they have $k_x=0$.

Now condense the basic dislocation in the y-direction, the
$|1\rangle$ defect. The phonons all braid nontrivially with the
condensate! Just as above, all the phonons are confined.

Now that we've seen how the symmetry is restored, we can also
discuss what happens in the Abrikosov lattice in Type II
superconductors, for example, in our terminology. The symmetry is first
fully restored by the defect condensate, and then it is broken to a lattice,
 as it were by the condensation of two defect density modes, similar
to the formation of the smectic. It is a dual version of what we
just discussed.

\section{Non-Abelian defect condensates }

We now want to move on to defect condensates in a non-Abelian
setting, and it turns out that we then have to distinguish different
types of defect condensates. Consider a phase described by a quantum
double $D(H)$, or a generalized quantum double $F(H_m)\otimes\C
H_{el}$\footnote{In this paper we adopt a notation with a tensor
product symbol "$\otimes$" between the magnetic and electric part of
the double, to keep the notation unambiguous in the cases we are to
investigate. The notation used in other papers we refer to is the
ordinary product, but in this paper we have to deal with electric
and magnetic symmetries which themselves involve a number of
products.}, and pick a purely magnetic representation $\Pi^A_{1}$
($1$ is the trivial representation of the centralizer $N_A$). A
basis of the vector space on which this irrep acts is given by
$\{|g^A_i\rangle\}$, where the $g^A_i$ are the different defects in
the conjugacy class \footnote{$C_A$ is a conjugacy class in the
$D(H)$ case, and an orbit in $H_m$ under the action of $H_{el}$ in
the \mbox{$F(H_m)\otimes\C H_{el}$} case.} $C_A$ whose
representative element we denote by $g_A$. In that defect class
$C_A$ we may consider the following inequivalent types of
condensates:
\begin{itemize}
\item Single defect condensate
\begin{equation}
\phi=|g^A_i\rangle
\end{equation}
\item Class sum defect condensate
\begin{equation}
\phi=\sum_{g^A_i \in A} |g^A_i\rangle =: |C_A\rangle.
\end{equation}
We denote the condensate by $|C_A\rangle$, where $g_A$ is the
preferred element of $\Acal$\footnote{We note that this condensate
is in fact gauge invariant (under $H$ transformations). This means
that if we are discussing a local theory these condensates would
also be acceptable ground states. For the case of global
symmetries such restrictions are not necessary.}.
\item Combined defect condensate
\begin{equation}
\phi=\sum_{g_i\in E} |g_i\rangle
\end{equation}
where $E$ is some subset of the defects in one class. We need only
take the elements to be within one class because it turns out that
we need only study the cases where the condensate is the sum of
vectors in the same irrep.
\end{itemize}
The single defect and class sum defect condensates are a special
case of the combined defect condensate. In the following we will
encounter both class sum and combined defect condensates.

\subsection{The hexagonal crystal and its regular modes}

We start with the planar, achiral, hexagonal crystal which has a
space symmetry group $\Z^2\sd\Z_6$.   Strictly speaking in this case
the defect classes (i.e. the sets of defects transformed into each
other under residual global symmetry transformations) are larger
than the conjugacy classes of the magnetic group, but for our
present purposes it suffices
to restrict our considerations to the conjugacy classes.\\
\begin{figure}[htb]
\begin{center}
\vspace{1in}
\includegraphics[scale=0.3,angle=270]{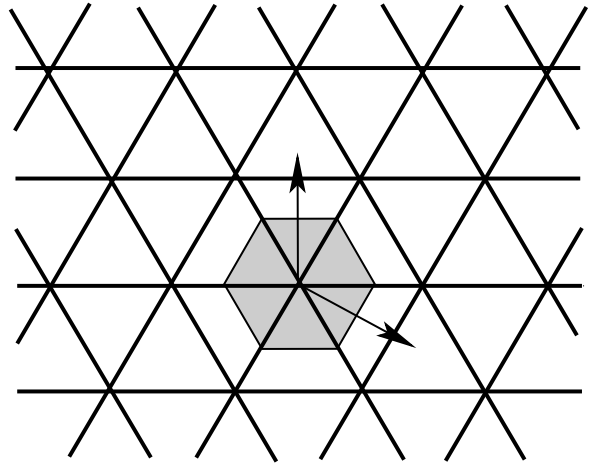}
\vspace{0.5in} \caption{A triangular lattice, representing a crystal
with hexagonal symmetry. The dashed vectors are vectors in the
reciprocal lattice, and the shaded hexagon represents the first
Brillouin zone.} \label{fig:FBZ}
\end{center}
\end{figure}
The first Brillouin zone (rescaled) is shown in Fig.\ref{fig:FBZ},
it is a hexagon. Points in the Brillouin zone correspond to irreps
of the translation group. Namely, a point in the Brillouin zone is a
momentum vector $\vec{k}=(k_x,k_y)$, and the associated irrep of
$\Z^2$ is:
\begin{equation}
(n,m)\mapsto e^{i (k_xn + k_ym)}.
\end{equation}

\subsection{Topological defects in the hexagonal phase}

The homotopy groups which define the defects are:
\begin{eqnarray}
&&\Pi_1(G/H)=\widetilde{H}\\
&&\Pi_2(G/H)= 0,
\end{eqnarray}
where $\widetilde{H}$ is the lift of $H$ in the universal covering group
$\widetilde{G}$ of $G$.
In three dimensions, there are no monopoles because the space
group $H$ is discrete. The line defects are
characterized by elements of $\widetilde{H}$.\\
First we consider a two-dimensional hexagonal crystal. To determine
$\widetilde{H}$ we need to lift $H$ to the covering group of
$G=\R^2\sd SO(2)$. The translational part $\R^2$ is already simply
connected. The covering group of $SO(2)$ is $\R$. An element $\te$
of $\R$ corresponds to a rotation over an angle $\te$. In the
covering group, however, a rotation over $2\pi$ is no longer
equivalent to the identity. Similarly, the covering of $\Z_6$ is
\begin{equation}
\widetilde{\Z_6}=\Z\times\Z_6\simeq\Z
\end{equation}
Therefore $\widetilde{\Z^2\sd\Z_6}\simeq\Z^2\sd\Z$, and
\begin{equation}
\Pi_1(G/H)\simeq\Z^2\sd\Z. \label{defecthomotopy}
\end{equation}
Therefore a defect is an element of $\Z^2\sd\Z$. We denote it by
$(n,m)r^p$, with $n,m,p\in \Z$. $r$ is a $\frac{2\pi}{6}$ rotation,
(1,0) a translation by one lattice vector in the x-direction, and
(0,1) a translation by one lattice vector in the y-direction. A
defect of the form $(n,m)$ (i.e. with $p=0$) is a dislocation. An
$r^p$ defect ($n=m=0$) is a disclination. Note that the structure of
defects described by (\ref{defecthomotopy}) is indeed non-Abelian,
which means that we have to think of them forming multiplets
corresponding to the conjugacy classes of the first homotopy group.
We return to this question in the following section when we
construct the Hopf algebra describing this phase.

Note that although both rotational and translational Goldstone modes
are present in a two-dimensional crystal, it is known that the
rotational Goldstone mode, called the \emph{roton}, is much more
massive\cite{Nelsonart} than the translational modes, because it
costs a lot of energy to rotate different parts of the crystal
relative to each other. For the same reason, \emph{disclinations are
much more massive than dislocations}, and at low temperatures we
only have dislocations and translational Goldstone modes.

For a three-dimensional crystal, the analysis is very similar. For
example, in a crystal made up of layers of hexagonal crystal,
$G=\R^3\sd SO(3)$ is broken to $H=\Z^3\sd \Z_6$, so the defects have
an extra label, which represents translational defects in the
direction perpendicular to the layers:
\begin{equation}
\Pi_1(G/H)=\Z^3\sd\Z
\end{equation}
For a dislocation characterized by $(n,m,k)$, the vector $(n,m,k)$
is the Burgers vector of the defect. A general dislocation has both
screw and edge character, and the Burgers vector is given by the
integral of the order parameter field $\vec{u}(\vec{x})$
along a loop L encircling the core\cite{Chaikin}. \\
The fusion rules of line defects have some interesting features,
which are crucial for the story to come. We illustrate these
features for the two-dimensional hexagonal crystal. To determine
the fusion rules, we need the conjugacy classes of the defects.
This requires knowledge of the multiplication in $\Z^2\sd\Z_6$,
which is set by
\begin{equation}
r (a,b) r^{-1}=(-b,a+b).
\end{equation}
The conjugacy classes are listed in
Table \ref{tab:hexcrysdef}.
\begin{table}[t!]
\begin{center}
\begin{tabular}{|c|c|}
\hline & \\ Representative $g_A$&
Conjugacy class  $C_A$               \\
& \\ \hline \hline & \\$r^{6k},\;k\in\Z$ &
$C_{r^{6k}}=\{r^{6k}\}$                     \\
& \\

$(a,b)r^{6k}$ &
$C_{(a,b)r^{6k}}= $\\
$k\in\Z,\; a,b \geq 0$& $ \{(a,b)r^{6k},(-b,a+b)r^{6k}, $\\
& $ (-a-b,a)r^{6k}, (-a,-b)r^{6k},$      \\
& $(b,-a-b)r^{6k},  (a+b,-a)r^{6k}\}$                    \\
& \\

$r^{1+6k},\;k\in\Z$ &
$C_{r^{1+6k}}= $\\
& $ \{(m,n)r^{1+6k} : (m,n)\in\Z^2 \}$                    \\& \\
$r^{-1+6k},\;k\in\Z$ &
$C_{r^{-1+6k}}= $\\
& $ \{(m,n)r^{-1+6k} : (m,n)\in\Z^2 \}$                  \\& \\
 $r^{2+6k},\;k\in\Z$ &
$C_{r^{2+6k}}= $\\
& $ \{(m,n)r^{2+6k} : m-n\in 3\Z \}$                      \\& \\
 $(1,0)r^{2+6k},\;k\in\Z$ &
$C_{(1,0)r^{2+6k}}= $\\
& $ \{(m,n)r^{2+6k} : m-n\notin 3\Z \}$          \\& \\
$r^{3+6k},\;k\in\Z$ &
$C_{r^{3+6k}}= $\\
& $ \{(m,n)r^{3+6k} : (m,n)\in 2\Z\times 2\Z \}$          \\& \\
 $r^{3+6k},\;k\in\Z$ &
$C_{(1,0)r^{3+6k}}= $\\
& $ \{(m,n)r^{3+6k} : (m,n)\notin 2\Z\times 2\Z \}$      \\& \\
 $r^{4+6k},\;k\in\Z$ &
$C_{r^{4+6k}}= $\\
& $ \{(m,n)r^{4+6k} : m-n\in 3\Z \}$                      \\& \\
$(1,0)r^{4+6k},\;k\in\Z$ &
$C_{(1,0)r^{4+6k}}= $\\
& $ \{(m,n)r^{4+6k} : m-n\notin 3\Z \}$              \\& \\
\hline
\end{tabular}
\caption{The classification of defects for the 2-dimensional
hexagonal crystal in classes of
$\overline{\Z^2\sd\Z_6}=\Z^2\sd\Z$.} \label{tab:hexcrysdef}
\end{center}
\end{table}
From this we can for example prove that two disclinations can fuse
to form any dislocation in this crystal. Namely, the fusion rules
are given by multiplying entire conjugacy classes. If we fuse $C_r$
and $C_{r^{-1}}$, we can get any dislocation $(m,n)$ because
$(m,n)r$ is an element of $C_r$. Thus, from a topological point of
view a dislocation can ``dissociate'' into two disclinations.\\

\subsection{The double symmetry of the hexagonal crystal}

In the previous subsection we saw that the magnetic group of a
hexagonal crystal with large symmetry transformations is
$\Z^2\sd\Z$. Let us now consider a hexagonal crystal and include
the reflections, then the magnetic group is the same\footnote{The
inversions are not connected to the identity, therefore adding
them does not influence $\Pi_1(G/H)$.}: $H_m=\Z^2\sd\Z$. The
electric group is $H_{el}=\widetilde{\Z^2\sd D_6}=(\Z^2\sd
D_6)\times\Z$. The last $\Z$ part corresponds to rotations of
$2\pi n$, $n\in \Z$, which we denote by $r^{6n}$. These rotations
are in the center of the group. Thus the Hopf symmetry
is\footnote{There is a subtlety due to the infinity of the groups.
Drawing on the analogy with generalized quantum doubles with
finite groups, as a vector space this Hopf algebra is \mbox{$F(
((\Z^2\sd \Z_6)\times \Z)) \tensor \C (\Z^2\sd D_6))$}. However, to
make the representation theory tractable we only take functions
with compact support in the second variable. One can check that
this is an algebra (i.e. the finite sum and product of such
elements also has compact support in the second variable). We
don't take compact support in the first variable because of our
analysis of defect condensates, to come. If we condense the
dislocation $|(1,0)\rangle$, for example, then we need functions
that are constant on left $(1,0)$ cosets, and these left cosets
have an infinite number of elements.}:
\begin{equation}\label{hexagonalhopf}
\mathcal{A} = F(\Z^2\sd\Z)\otimes\C ((\Z^2\sd D_6)\times\Z).
\end{equation}

We write an element of $(\Z^2\sd D_6)\times\Z$ as $(k,l)s^j r^{k+6n}$,
with $k,l,n\in\Z$, $j\in\Z_2$, $n\in\Z$. The reflection in $D_6$
is denoted by $s$. We take $s$ to be the reflection
with respect to the x-axis.\\
To analyze the quantum numbers of the excitation sectors of our
theory, we proceed as usual and have to construct the
representations of the algebra given
above\cite{Bais:2002pb,Bais:2002ny,bmbreaking:2006}. The
representations of the generalized quantum doubles have a generic
structure. There is an orbit or defect class of the group in the
function part labelling the defects, paired with a representation of
the centralizer $N_A$ for the ordinary/Goldstone modes. These
representations will be constructed in this section and the results
are listed in Table \ref{tab:hexdouble}. We need the orbits $\Orb$
of $((\Z^2\sd \Z_6)\times \Z)$ under the action of $(\Z^2\sd
D_6)\times\Z$, the centralizer $N$ of a chosen preferred element of
each orbit, and the irreps of the centralizer. The action of
$(\Z^2\sd D_6)\times\Z$ on $\Z^2\sd \Z_6$ is given by:
\begin{eqnarray}
&&r(a,b)r^{-1}=(-b,a+b),\nonumber\\
&&s(a,b)s^{-1}=(a+b,-b).\nonumber\\
&&srs^{-1}=r^{-1}\nonumber
\end{eqnarray}

\begin{table}[t!]
\centering
\begin{tabular}{|c|}
\hline \\
Orbits \\  \\
\hline\hline  \\
$\Orb_{r^{6k}}= C_{r^{6k}}$\\[2mm]
$\Orb_{(a,b)r^{6k}}= \{C_{(a,b)r^{6k}},C_{(a+b,-b)r^{6k}}\}$\\[2mm]
$\Orb_{r^{1+6k}}= \{C_{r^{1+6k}},C_{r^{-1-6k}}\}$\\[2mm]
$\Orb_{r^{2+6k}}= \{C_{r^{2+6k}},C_{r^{-2-6k}}\}$\\[2mm]
$\Orb_{(1,0)r^{2+6k}}= \{C_{(1,0)r^{2+6k}},C_{(1,0)r^{-2-6k}}\}$\\[2mm]
$\Orb_{r^{3+6k}}= \{C_{r^{3+6k}},C_{r^{-3-6k}}\}$\\[2mm]
$\Orb_{(1,0)r^{3+6k}}=
\{C_{(1,0)r^{3+6k}},C_{(1,0)r^{-3-6k}}\}$\\\mbox{}\\[2mm]
 \hline
\end{tabular}
\caption{The orbits of $\Z^2\times\Z$ under the action of
$(\Z^2\sd D_6)\times\Z$. The conjugacy classes are defined in
Table \ref{tab:hexcrysdef}.} \label{tab:crysclas}
\end{table}

\begin{table}[t!]
\begin{center}
\begin{tabular}{|c|c|c|}
\hline & & \\
Conjugacy classes           &       $N$         & Irreps $\Pi^A_\alpha$\\ & & \\
\hline\hline & & \\
$C_{r^{6k}}$            &   $(\Z^2\sd D_6)\times\Z$ &
$\Pi^{r^{6k}}_{\rho,\lambda}$     \\
$k\in\Z$                    &       & \\
 & & \\
$C_{(a,b)r^{6k}},C_{(a+b,-b)r^{6k}}$&$(\Z^2)\times\Z$
&$\Pi^{(a,b)r^{6k}}_{\vec{k},\lambda}$\\
$a,b\geq 0; b\geq a; k\in \Z$ &       &$\vec{k}\in\R^2$       \\ & & \\

$C_{r^{1+6k}},C_{r^{-1-6k}}$    &   $(\Z_6)\times\Z$    &
$\Pi^{r^{1+6k}}_{m,\lambda}$    \\
$k\in\Z$                            &           &   $m\in\Z_6$          \\ & & \\

$C_{r^{2+6k}},C_{r^{-2-6k}}$    &   $(\Z_6)\times\Z$    &
$\Pi^{r^{2+6k}}_{m,\lambda}$ \\
        $k\in\Z$            &                   &       $m\in\Z_6$          \\ & & \\

$C_{(1,0)r^{2+6k}},C_{(1,0)r^{-2-6k}}$&$(\Z_3)\times\Z$
&$\Pi^{(1,0)r^{2+6k}}_{n,\lambda}$\\
        $k\in\Z$                                    &           &   $n\in\Z_3$  \\ & & \\

$C_{r^{3+6k}},C_{r^{-3-6k}}$    &$(\Z_6)\times\Z$   &   $\Pi^{r^{3+6k}}_{m}$    \\
        $k\in\N$                &                   &       $m\in\Z_6$          \\ & & \\

$C_{(1,0)r^{3+6k}},C_{(1,0)r^{-3-6k}}$
&$(\Z_2)\times\Z$&$\Pi^{(1,0)r^{3+6k}}_{l,\lambda}$\\
            $k\in\N$                            &           &   $l\in\Z_2$  \\ & & \\
\hline
\end{tabular}
\caption{The irreps of $F( (\Z^2\sd\Z )\otimes\C(\Z^2\sd
D_6)\times\Z$. The index $\lambda\in [0,2\pi)$ corresponds to the
irreps of the $\Z$ part of the centralizers, which represents the
rotations by an angle $2\pi n$. The index $\rho$ in the first
representation listed, refers to the irreps of $\Z^2\sd D_6$
defined in Table \ref{tab:hexirreps}.} \label{tab:hexdouble}
\end{center}
\end{table}

There is one class of orbits, $\Orb_{r^{6k}}$, for which
determining the irreps of the centralizer $(\Z^2\sd D_6)\times\Z$
is a little complicated. We can determine the irreps of $\Z^2\sd
D_6$ and those of $\Z$ separately, because the irreps of a direct
product of groups are the tensor products of the irreps of the
separate groups. The irreps of $\Z^2\sd D_6$ can be constructed by
the method of induced representations described in the appendix.
We must pick a momentum vector $\vec{k}$ which corresponds to an
irrep of the translation group $\Z^2$, and act on it with $D_6$.
The elements that transform $\vec{k}$ into a vector that
corresponds to the same irrep form a subgroup of $D_6$, which we
denote by $(D_6)_{\vec{k}}$. Thus we are considering orbits in the
first Brillouin zone (note that these classes have nothing to do
with the orbits listed in Table \ref{tab:hexcrysdef}). We must
find all possible cases of vectors in the first Brillouin zone,
with different $(D_6)_{\vec{k}}$. For the hexagonal crystal, there
are seven different types of orbits, and a representative element
of each type is given in Figure \ref{fig:hexagonal}.
\begin{figure}[htb]
\centering
\includegraphics[scale=0.2]{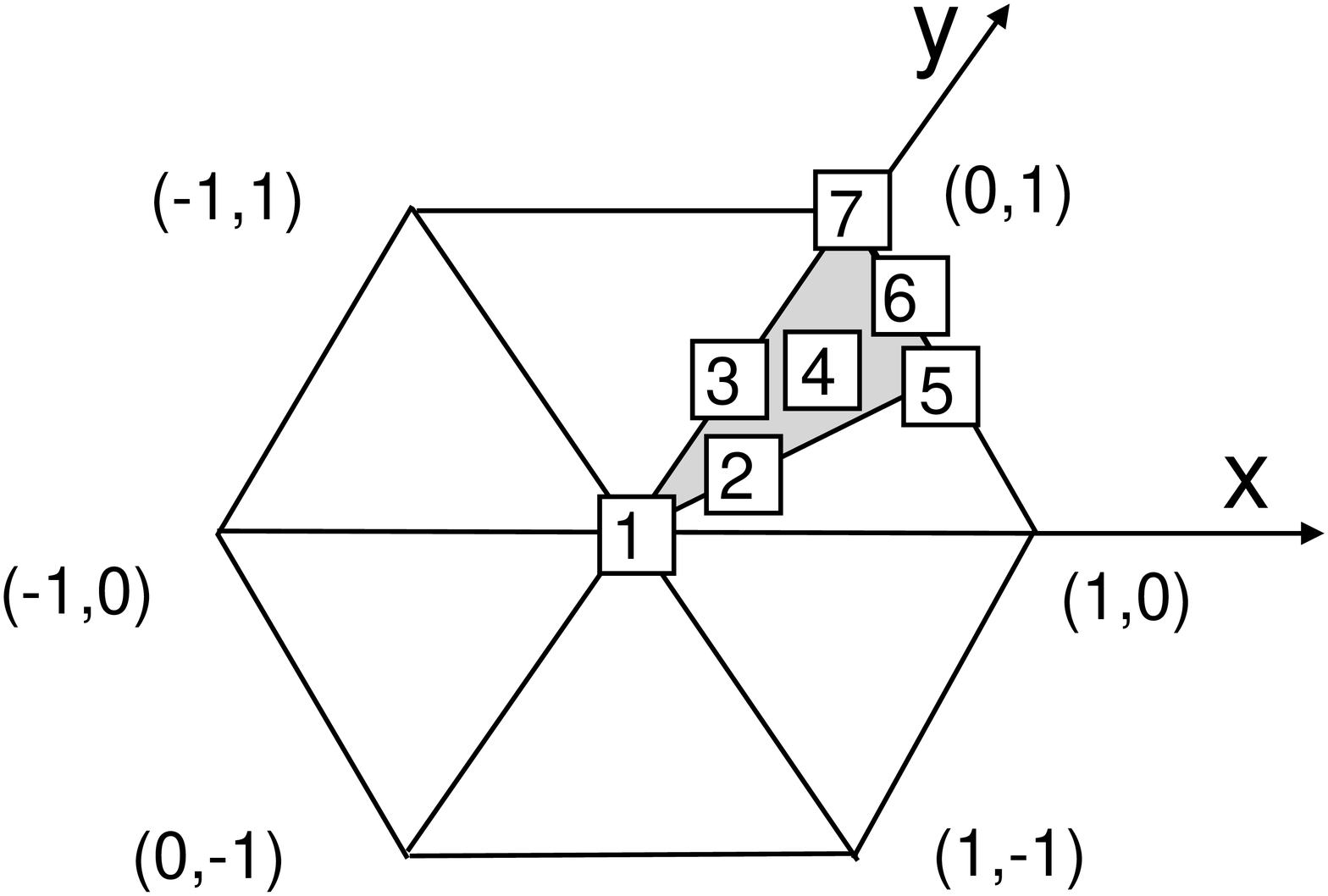}
\caption{The first Brillouin zone of the planar hexagonal crystal.
The gray area is a set of representatives of orbits under
rotations and reflections. The numbered squares are
representatives of the different types of orbits. The x- and
y-axes have been rescaled. The unit distance on these rescaled
axes corresponds to a distance of $\frac{\pi}{a}$, where $a$ is
the lattice spacing in the crystal.} \label{fig:hexagonal}
\end{figure}

\begin{table}[h!]
\begin{center}
\begin{tabular}{|c|c|c|c|}
\hline & & & \\
Orbit number    &   $(D_6)_{\vec{k}}$&  Number of irreps    &   Dim. of irreps \\ & & & \\
\hline \hline & & & \\
1               &   $D_6$           &   6                   &   1, 1, 1, 1, 2, 2    \\
2               &   $\Z_2$          &   2                   &   6, 6                \\
3               &   $\Z_2$          &   2                   &   6, 6                \\
4               &   $e$             &   1                   &   12                  \\
5               &   $\Z_2\sd\Z_2$   &   4                   &   3, 3, 3, 3          \\
6               &   $\Z_2$          &   2                   &   6, 6                \\
7               &   $D_3$           &   3                   &   2, 2, 4             \\ & & & \\
\hline
\end{tabular}
\caption{The irreps of $\Z^2\sd D_6$. The number of an orbit
corresponds to the number in Fig. \ref{fig:hexagonal}.
$(D_6)_{\vec{k}}$ is the Little Group of the orbit.}
\label{tab:hexirreps}
\end{center}
\end{table}

We then call $\Z^2\sd(D_6)_{\vec{k}}$ the \emph{little group} of
$\vec{k}$. We must determine the irreps of the little group, and
then \emph{induce} irreps of the whole group. This procedure
gives all the irreps. which are given in Table \ref{tab:hexirreps}.\\
As an example, We explicitly give the irrep corresponding to orbit
5 in Figure \ref{fig:hexagonal}. In the appendix it is illustrated
how to derive this. The matrices of the irrep are
\begin{displaymath} (a,b) \mapsto \left( \begin{array}{ccc}
            e^{i \frac{\pi}{2} a} e^{i \frac{\pi}{2} b}     & 0     & 0 \\
            0   & e^{i \frac{\pi}{2} (a+b)} e^{i \frac{\pi}{2} (-a)} & 0  \\
            0           & 0         & e^{i \frac{\pi}{2}b} e^{i \frac{\pi}{2}(-a-b)}
            \end{array} \right),
\end{displaymath}
\begin{displaymath} r \mapsto \left( \begin{array}{cccc}
            0 & 0 & e^{i \pi m} \\
            1 & 0 & 0 \\
            0 & 1 & 0
            \end{array} \right),
            s \mapsto \left( \begin{array}{cccc}
            0           &               & e^{i\pi n}\\
            0           & e^{i\pi n}    & 0         \\
            e^{i\pi n}  & 0             & 0
            \end{array} \right)
\end{displaymath}

\subsection{Non-Abelian defect condensates in hexagonal crystals}
Let us now turn to the \textit{hexatic} phase. Before turning to
the Hopf symmetries involved in the next section, we first recall
some of the standard analysis.  The phase is two dimensional. It
is translationally invariant, but the rotational group is broken
to $\mathbb{Z}_6$. Such a phase looks strange at first sight,
because it seems that to break rotational symmetry we need to put
the atoms on a lattice, thereby also breaking translational
symmetry. We know that this is not the case because such a hexatic
phase is actually realized in nature \cite{Mullen,Nelson}.

That it is \emph{not} impossible to break translational symmetry
without breaking rotational symmetry, can be understood from the
representation theory\footnote{If an irrep is trivial when
restricted to $\R^2$, when restricted to $SO(2)$ it can be any
irrep of $SO(2)$. If, on the other hand, it is trivial on $SO(2)$,
then the irrep must act the same on all vectors of the same
length, since we can turn any two unit vectors into each other
with a transformation of $SO(2)$. This leads to the conclusion
that the irrep is trivial on $\R^2$.} of $\R^2\sd SO(2)$. A
suitable order parameter field is provided by $\phi=e^{i6\te}$,
where $\te$ measures the angle the bonds between neighboring
molecules makes with the x-axis. In the high-temperature isotropic
phase, $<e^{i6\te}>=0$. In the ordered phase, $<e^{i6\te}>\neq 0$
because the angle of the bonds is on average a multiple of
$\frac{2\pi}{6}$.

The hexatic phase can arise via spontaneous symmetry breaking from
the isotropic phase, but it can also be the outcome of the
\emph{melting} of a two-dimensional hexagonal crystal. This is
what happens in nature: as we increase the temperature of the
crystal, at some point the translational symmetry is restored, but
locally the molecules are still arranged hexagonally. Of course,
locally there is some translational symmetry, but the
correlations of translations decay exponentially. \\
We noted before that in a two-dimensional crystal the disclinations
and the roton are very massive, and are not excited at low
temperatures. As the temperature is increased and the translational
symmetry is \emph{restored} (signalling the transition to the
hexatic phase), the translational Goldstone modes and the
dislocations disappear from the spectrum. At the same time, the
roton becomes massless, while the disclinations are the allowed
defects in this phase.\\
The traditional phase diagram of a two-dimensional crystal
contains a line separating the crystal phase and the liquid phase.
The phase transition is first order. However, for the hexagonal
crystal the transition to the liquid phase can occur via two
second-order phase transitions: first a phase transition to the
translationally invariant phase with broken rotational symmetry,
and then a transition to the liquid.

The mechanism behind the phase transition from crystal to hexatic is
well known, and quite remarkable: it is the formation of a
\emph{dislocation condensate} \cite{Nelsonart}. Defects in a phase
with global symmetry breaking cost an infinite amount of energy,
because their energy diverges logarithmically with the size of the
system\cite{Chaikin}. However, defect-antidefect pairs cost a finite
amount of energy, and at finite temperatures they have a finite
density, and an average separation between the defects in a pair,
which we call $a$. At the phase transition, $a$ diverges, so that
the single defects fill the ground state. This \emph{restores
translational symmetry}. In the next section we will discuss why it
restores symmetries. Here we merely note what happens next if the
ground state is now filled with single dislocations. We've seen that
these dislocations in principle can decay into a
disclination-antidisclination pair. They also have an average
separation, $b$. At the second phase transition, the disclinations
blow out, thereby \emph{restoring rotational symmetry}.

The hexatic phase has been experimentally observed in numerous three
dimensional phases, such as smectics, dense solutions of DNA, flux
arrays of High-$T_c$ superconductors, superfluids \cite{Mullen},
etc. All these phases have some kind of layered order. The hexatic
smectic is also called the smectic B phase. The bond-orientational
order is within the layers of the smectic. For references on hexatic
phases, we refer to the book by Nelson\cite{Nelson}.

\subsection{The hexatic phase}

Let us now apply our methods to this situation of the hexatic phase
as an example of defect mediated melting. We start with the Hopf
symmetry of the hexagonal crystal:
\begin{equation}
\mathcal{A}=F(\Z^2\rtimes(\Z_6\times\Z))\otimes\C((\Z^2\rtimes
D_6)\times\Z),
\end{equation}
and its representations which were given in Table
\ref{tab:hexdouble}.

 Now we consider a class sum defect condensate of the basic
 dislocation $(1,0)$ belonging to the class $C_{(a,b)r^{6k}}$ with $k=0$ (see
Figure \ref{fig:hexdef}):
\begin{eqnarray}\label{hexcondensate}
|\phi_0\rangle&=& |(1,0)\rangle +|(0,1)\rangle+|(-1,1)\rangle \nonumber\\
&&+|(-1,0)\rangle+|(0,-1)\rangle+|(1,-1)\rangle.
\end{eqnarray}
It is not hard to determine what the intermediate $\Rcal$ and
unconfined $\mathcal{U}$ symmetries are in this case (the derivation
is carried out in a related paper\cite{bmbreaking:2006}):
\begin{eqnarray}
&&\Rcal = F(\Z^2\rtimes(\Z_6\times\Z)/\Z^2)
\otimes\C((\Z^2\rtimes D_6)\times\Z) \nonumber\\
&&\mathcal{U} = F(\Z_6\times\Z)\otimes\C (D_6\times\Z).
\end{eqnarray}
$\Z_6\times\Z$ corresponds to disclinations, and $D_6\times\Z$
corresponds to rotations and inversions. The $\Z$ part of
$D_6\times\Z$ corresponds to multiples of the $2\pi$ rotation.

\begin{figure}[h]
\centering
\includegraphics[scale=0.2]{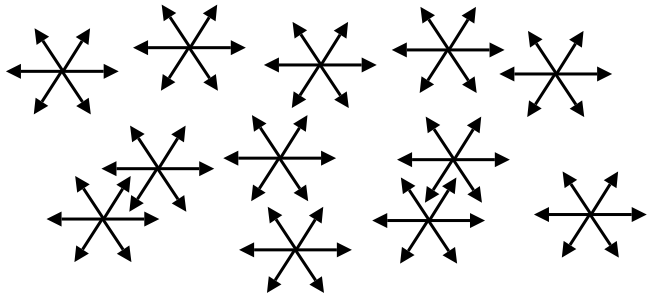}
\caption{An artist impression of the class sum dislocation defect
condensate $|\phi_0\rangle =
|(1,0)\rangle+|(0,1)\rangle+|(-1,1)\rangle+|(-1,0)\rangle+|(0,-1)\rangle+|(1,-1)\rangle$
with hexagonal symmetry.} \label{fig:hexdef}
\end{figure}

The particular condensate of dislocations is invariant under
rotations, hence disclinations are not confined. Basically the
conclusion of our analysis is that the translational modes of the
lattice have completely disappeared. There are no more
dislocations (because they are condensed), and the translational
phonons are confined. As we discussed earlier, this signals the
restoration of translational symmetries, so the unconfined and
residual symmetry algebras of the hexatic phase are actually
\begin{eqnarray}
&&\Rcal = F(\Z^2\rtimes(\Z_6\times\Z)/\Z^2)
\otimes\C((\R^2\rtimes D_6)\times\Z) \nonumber\\
&&\mathcal{U} = F(\Z_6\times\Z)\otimes\C (\R^2\rtimes
(D_6\times\Z)).\label{hexatichopf}
\end{eqnarray}

We have thus succeeded in interpreting the phase transition from
the hexagonal crystal to the hexatic phase (including also the
translational symmetry restoration), as a Hopf symmetry breaking
phenomenon through a particular defect class condensate.
\subsection{A topological nematic}
We now wish to comment on a phase  called the \emph{topological
nematic}\cite{Zaanen}. This is a two dimensional phase obtained from
an isotropic crystal, where dislocations have condensed, but
rotational symmetry is completely unbroken. In the defect condensate
(\ref{hexcondensate}) that led to the hexatic phase, the residual
rotational symmetry group was $\Z_6$, but in the topological nematic
the rotational symmetry group is $U(1)$. Clearly, an isotropic
crystal only makes sense in the continuum limit, because by
definition a crystal is not isotropic. However, in the continuum
limit one can definitely have a Lagrangian of a crystal in terms of
the displacement field $\vec{u}(\vec{x})$, that is invariant under
$ISO(2)=\R^2\rtimes SO(2)$. One can show that the two dimensional
hexagonal and isotropic crystals have exactly the same Lagrangians
\cite{Landau} (they only have two independent elastic constants), so
we could consider this isotropic crystal to be the continuum limit
of a hexagonal crystal.

Now the topological nematic is a phase obtained by a defect
condensate in the isotropic crystal. To obtain a defect condensate
that fully restores the rotational symmetry group $SO(2)$, we must
first realize that in the continuum limit the dislocations carry a
continuous label. The Burgers vector of a dislocation in the
continuum limit is characterized by a two-dimensional vector
$|(a,b)\rangle$, with $a, b \in \R^2$. Thus a dislocation
condensate that leads to restoration of the full rotational
symmetry group is given by
\begin{equation}
|\phi_r\rangle = \int_0^{2\pi} d\theta \;|(cos\theta,
sin\theta)\rangle.
\end{equation}
Strictly speaking, this condensate is the continuum limit of the
class sum defect condensate we considered in the hexatic phase. It
is indeed a class sum defect condensate of the isotropic crystal.
It is in that sense that our analysis naturally incorporates the
topological nematic.

Furthermore, our analysis demonstrates how  the partial
restoration of the symmetry group can be understood by considering
defect condensates that are not a class sum, as we will show next.
Let us return to the hexatic phase  and analyze the transition to
the isotropic phase. The hexatic phase is a phase described by the
Hopf algebra (\ref{hexatichopf})
\begin{equation}
\mathcal{A}=F(\Z_6\times\Z)\otimes\C (\R^2\rtimes
(D_6\times\Z)).\nonumber
\end{equation}
Now assume a disclination condensate of $|r\rangle$, the
$\frac{2\pi}{6}$ rotation, then the residual symmetries are
successively
\begin{eqnarray}
&&\Rcal = F(e)\otimes\C(\R^2\rtimes (\Z_6\times\Z)) \nonumber\\
&&\mathcal{U} = \C \R^2.
\end{eqnarray}

\subsection{A quantum smectic?}
We mentioned that with non-Abelian defects there are in general
quite a few  possibilities for the condensate. Our formalism allows
us to determine what the physical implications are. We conclude by
considering, instead of the class condensate (\ref{hexcondensate}),
the defect-antidefect condensate,
\begin{equation}
|\phi_0\rangle=|(1,0)\rangle+|(-1,0)\rangle.\nonumber\\
\end{equation}
The residual and unconfined symmetry algebras turn out to be
\begin{eqnarray}
&&\Rcal = F( (\Z^2\rtimes\Z_6)\times\Z) /{0}\times \Z)
\otimes \C (\Z^2 \rtimes (\Z_2\rtimes\Z_2)) \nonumber\\
&&\mathcal{U}=F((\Z\rtimes\Z_2)\times\Z)\otimes\C(\Z\rtimes
(\Z_2\rtimes\Z_2)).
\end{eqnarray}
The dislocations along the $(1,0)$ direction are condensed, and
the phonons with momentum in that direction are confined.
Consequently the symmetry is restored in that direction, so that
\begin{eqnarray}
&&\Rcal = F( (\Z^2\rtimes\Z_6)\times\Z) /{0}\times\Z) \otimes
\C (\Z^2 \rtimes (\Z_2\rtimes\Z_2)) \nonumber\\
&&\mathcal{U}=F((\Z\rtimes\Z_2)\times\Z)\otimes\C((\R\times\Z)
\rtimes(\Z_2\rtimes\Z_2)).
\end{eqnarray}

This is a 2D smectic. Note that in a conventional classical smectic,
the Goldstone modes corresponding to oscillations of the rods within
a plane are massive (because of the analog of the Higgs phenomenon
\cite{Chaikin}). Determination of whether the modes are massive or
not requires a dynamical analysis. What we do know from our
topological analysis is that these Goldstone modes are not confined.

The mass term for the Goldstone modes in classical smectics arises
because the internal symmetry of the rods is coupled to the
external symmetries (the Lagrangian is only invariant under
coupled internal and external rotations). In the 2D smectic we
have uncovered now, there are no internal symmetries. However, the
defect-antidefect pairs in our condensate behave like rods in a
conventional smectic. Note that they are coupled to external
symmetries, in fact they arise due to the breaking of external
symmetries. Thus we also expect the Goldstone modes associated
with the oscillation of these ``rods'' to be massive, and only the
longitudinal oscillation of the planes to be massless. And so our
analysis has naturally led us to the quantum smectic discussed by
Zaanen and Nussinov \cite{Zaanen}.

\section{Conclusions}

In this paper we have applied the formalism of quantum double
symmetry breaking to a number of physical phenomena linked to the
formation defect condensates. We showed
that both for cases with internal and external symmetries,
nontrivial known and conjectured results can be recovered. It is
especially fruitful to apply the method in cases where one is
dealing with non-Abelian symmetries and consequently with
non-Abelian defects, because in such situations many inequivalent
condensates can be considered, which lead to effective theories
with very different low energy degrees of freedom. It is also nice
that the criterion for confinement determines to what extent a
restoration of the original gauge symmetry takes place. The method
enables one also to systematically analyze dyonic condensates, but
in this paper we have restricted ourselves to pure defect
condensates.

It is clear that there are many two dimensional (quantum) systems
that can be systematically studied from this general point of view.
In a paper appearing in parallel we treat non-Abelian nematic
crystals with nontrivial point group symmetries of the tetrahedral,
octahedral and icosahedral type \cite{mbnematic:2006}. All systems
in which conformal field theory plays a role, such as quantum Hall
systems, or Bose-Einstein condensates, fall naturally  in the
category that should be analyzed. For example it would be
interesting to investigate wether the plateau hierarchy proposed by
Haldane \cite{Haldane:1983}, where the plateaus are linked by the
successive formation of quasi particle condensates, could possibly
fit in our scheme. It may also be interesting to investigate the
conceivable phases of (2+1)-dimensional gravity, along these lines.

 \noindent {\bf Acknowledgement:} One of us F.A.B., thanks the
Research School for the Physical Sciences and Engineering at the
Australian National University in Canberra, where part of this work
was carried out, for their stimulating hospitality.

\vspace{0.5in}

\section*{Appendix: \\Inducing representations of $\Z^2\sd D_6$}
\label{subsec:induced}

Here, we illustrate the derivation of an irrep of $\Z^2\sd D_6$
discussed in the main text, using Mackey's induction theorem (see
any good book on representation theory\cite{Serre}). Take orbit nr 5
in fig. \ref{fig:hexagonal}, with momentum
$\vec{k}=(\frac{\pi}{2},\frac{\pi}{2})$ (i.e.,
$(\frac{\pi}{2a},\frac{\pi}{2a})$ with $a=1$). A rotation of 180
degrees ($r^3$) gives an equivalent vector, i.e. a momentum that
represents the same irrep of $\Z^2$. As a matter of fact, so do $e,
sr^2, sr^{-1},$ so $H_{\Pi_{\vec{k}}}$ is isomorphic to
$\Z^2\sd\Z_2=(r^3)\times(sr^2)$, where in general $(g)$ is the group
generated by $g$. It has four one-dimensional irreps. Now
$[D_6:\Z^2\sd\Z_2]=3$, so we get four three-dimensional irreps.
Denote an irrep of $\Z^2\sd\Z_2$ by $\rho_{m,n}$, with $m,n\in\Z_2$.
Explicitly, a basis of the irreps is given by a set of
representatives of left $\Z^2\sd\Z_2$ cosets, which we take to be
$\{|e\rangle,|r\rangle,|r^2\rangle\}$ (the order is important when
we write down the matrices explicitly). The matrices of the irrep
are
\begin{displaymath} (a,b) \mapsto \left( \begin{array}{ccc}
            e^{i \frac{\pi}{2} a} e^{i \frac{\pi}{2} b}     & 0     & 0 \\
            0   & e^{i \frac{\pi}{2} (a+b)} e^{i \frac{\pi}{2} (-a)} & 0  \\
            0           & 0         & e^{i \frac{\pi}{2}b} e^{i \frac{\pi}{2}(-a-b)}
            \end{array} \right),
\end{displaymath}
\begin{displaymath} r \mapsto \left( \begin{array}{cccc}
            0 & 0 & e^{i \pi m} \\
            1 & 0 & 0 \\
            0 & 1 & 0
            \end{array} \right),
            s \mapsto \left( \begin{array}{cccc}
            0           &               & e^{i\pi n}\\
            0           & e^{i\pi n}    & 0         \\
            e^{i\pi n}  & 0             & 0
            \end{array} \right)
\end{displaymath}
As a representative example, let us illustrate the derivation of the
second column of the matrix corresponding to s (the derivations of
the other columns is analogous):
\begin{equation}
s|r\rangle=|sr\rangle=|r(sr^2)\rangle=|r\rangle e^{i\pi n}.
\end{equation}

\pagebreak
\bibliography{blob}
\bibliographystyle{unsrt}

\end{document}